\documentclass[10pt,conference]{IEEEtran}
\IEEEoverridecommandlockouts
\usepackage{cite}
\usepackage{amsmath,amssymb,amsfonts}
\usepackage[ruled,norelsize,vlined]{algorithm2e}
\SetKwRepeat{Do}{do}{while}%
\makeatletter
\newcommand{\removelatexerror}{\let\@latex@error\@gobble}
\makeatother
\usepackage[caption=false,font=footnotesize]{subfig}
\usepackage{array}
\usepackage{multirow}
\usepackage{booktabs}
\usepackage{graphicx}
\usepackage{textcomp}
\usepackage{xcolor}
\def\BibTeX{{\rm B\kern-.05em{\sc i\kern-.025em b}\kern-.08em
    T\kern-.1667em\lower.7ex\hbox{E}\kern-.125emX}}
\makeatletter
\newcommand{\linebreakand}{%
  \end{@IEEEauthorhalign}
  \hfill\mbox{}\par
  \mbox{}\hfill\begin{@IEEEauthorhalign}
}
\makeatother
\begin{document}

\title{AFETM: Adaptive Function Execution Trace Monitoring for Fault Diagnosis
\thanks{This work was supported in part by Alibaba Co. Ltd.

Jianhui Jiang is the corresponding author.}
}

\author{\IEEEauthorblockN{Wei Zhang}
\IEEEauthorblockA{\textit{Tongji University}\\
Shanghai, China \\
1910134@tongji.edu.cn}
\and
\IEEEauthorblockN{Yuxi Hu}
\IEEEauthorblockA{\textit{Alibaba Inc.}\\
Beijing, China \\
huyuxi.pt@alibaba-inc.com}
\and
\IEEEauthorblockN{Bolong Tan}
\IEEEauthorblockA{\textit{Alibaba Inc.}\\
Beijing, China \\
bolong.tbl@alibaba-inc.com}
\linebreakand
\IEEEauthorblockN{Xiaohai Shi}
\IEEEauthorblockA{\textit{Alibaba Inc.}\\
Beijing, China \\
xiaohai.sxh@alibaba-inc.com}
\and
\IEEEauthorblockN{Jianhui Jiang*}
\IEEEauthorblockA{\textit{Tongji University}\\
Shanghai, China \\
jhjiang@tongji.edu.cn}
}

\maketitle

\begin{abstract}
The high tracking overhead, the amount of up-front effort required to selecting the trace points, and the lack of effective data analysis model are the significant barriers to the adoption of intra-component tracking for fault diagnosis today. This paper introduces a novel method for fault diagnosis by combining adaptive function level dynamic tracking, target fault injection, and graph convolutional network. In order to implement this method, we introduce techniques for (i) selecting function level trace points, (ii) constructing approximate function call tree of program when using adaptive tracking, and (iii) constructing graph convolutional network with fault injection campaign. We evaluate our method using a web service benchmark composed of Redis, Nginx, Httpd, and SQlite. The experimental results show that this method outperforms log based method, full tracking method, and Gaussian influence method in the accuracy of fault diagnosis, overhead, and performance impact on the diagnosis target.
\end{abstract}


\section{Introduction}
Software log is usually the most used data for developers and maintainers to analyze the system states. When errors/failures occur, they usually query recent log and try to find useful information about errors/failures. However, software log is generally set subjectively by developers based on personal experience during development. The content, quantity, format and structure of log can not usually be changed in the system runtime \cite{log20}. Moreover, log may not contain any message about the errors/failures, or the recorded message can not directly indicate the root causes. The fault coverage of log for different software is quite different, some are even less than 1\% \cite{target_fault_injection}. In order to solve the above problems, software dynamic tracking technology, as a supplement to native logging mechanism, came into being.

The software dynamic tracking technology \cite{Dtrace} identifies and extracts the causal related events generated by the software system when processing the service request. It depicts the system behavior accurately in fine-grained, which is of great significance to improve the efficiency and accuracy of the software fault diagnosis. Software dynamic tracking technology can be divided into two categories: intra-component tracking \cite{dynamic_call_graph} and inter-component tracking \cite{Pinpoint}. Intra-component tracking reveals the local states and behavior of a software system. Inter-component tracking, also knowing as end-to-end tracking \cite{trace_application}, describes the execution logic of the component-based or distributed software system as a whole.

End-to-end tracking associates the component-level causal related events to generate the execution path of the request on each component. As the scale of distributed software systems becomes larger, the structure becomes more complex, and the degree of request concurrency becomes higher, end-to-end tracking technology has gradually become a research hotspot, which has attracted a lot of attention from academia and industry \cite{target_fault_injection, vPath, Dapper, Magpie, Xtrace, Canopy, Pivot, workflow_centric_tracing}. The huge commercial value has been proved by a large number of industrial applications \cite{trace_application}. However, the granularity of fault position found based on this type of method is relatively coarse (generally component-level fault), which leads to higher cost of maintenance. Long manual diagnosis time usually leads to large financial loss \cite{IDC}. Intra-component tracking obtains fine-grained runtime information, e.g., function-level execution path \cite{dynamic_call_graph}, which can be used to help maintainers to understand the system behavior more comprehensively. But unlike the end-to-end tracking, intra-component tracking is generally used in the software development and testing phases to help the program debugging \cite{trace_debugging}. The following three major problems limit the application of intra-component tracking in the production environment \cite{dynamic_call_graph,Dtrace,TAU}, and to the best of our knowledge, they have not been systematically resolved today.

1) \textbf{High tracking overhead}. For end-to-end tracking, it usually tracks few specific communication and process/thread management related functions, e.g., \textit{send}(), \textit{receive}(), \textit{fork}() \cite{vPath}, to construct the request execution path on each component. For intra-component tracking, It requires tracking more functions inside a component to depict the function-level execution path. The traditional intra-component tracking is mainly used for program debugging \cite{trace_debugging}, and little attention were payed to control the overhead of tracking \cite{dynamic_call_graph,TAU}. Tracking all the functions defined in a component usually introduces large overhead \cite{TAU}, especially for high-performance components.

2) \textbf{Manual selection of trace points}. Adjusting trace points according to actual needs can help us obtain more useful data and control the tracking overhead \cite{TAU}. However, the trace points selection is often subtle for intra-component tracking, and developers usually spend significant amounts their valuable time selecting them before the full benefit, e.g., larger structure coverage and lower tracking overhead, can be obtained \cite{Canopy}. The amount of up-front effort required to instrument systems is the most significant barrier to the adoption of intra-component tracking \cite{workflow_centric_tracing}. 

3) \textbf{Lack of effective data analysis model for adaptive tracking}. As far as we know, there is no suitable data analysis method which performs well for fault diagnosis when the trace points are adjusted according to the runtime actual needs. In these types of scenarios, traditional function call sequence analysis method suffers from large accuracy loss. The Function Call Tree (FCT) data model \cite{lprof}, which represents calling relationships between functions, keeps more program structural information than sequence data. However, it can not even be established. Because unlike the full tracking method which tracks all the functions defined in the program, in the adaptive tracking scenarios, only the selected functions are tracked. The \textit{caller} and \textit{callee} data recorded by traditional method \cite{lprof} can not support the construction of the FCT in these scenarios. And for the error characteristic analysis based fault diagnosis \cite{target_fault_injection}, the fault matching time is usually linearly related to the size of the error characteristic database. Large database will lead to unacceptable fault diagnosis response time.

In order to address the above challenges, this paper proposes a novel fault diagnosis framework using Adaptive Function Execution Trace Monitoring (AFETM). The main contributions are as follows.

1) \textbf{Trace points adaptive selection approach}. To achieving lower tracking overhead and higher structural coverage, the trace points selection problem is transformed into a Minimum Weighted Set Coverage Problem (MWSCP) \cite{MWSCP} with upper bound total weight constraint. In which, each function covers several Intermediate Representation (IR) basic blocks, and the tracking of each function has a certain weight (also called cost). A Max-Min Ant System (MMAS) based runtime adaptive function-level trace points selection method has been proposed to solve this problem, so that the trace points can be generated adaptively with the goal of optimizing the tracking overhead and structural coverage comprehensively.

2) \textbf{Function execution trace data model}. A novel Approximate Function Call Tree (AFCT) data model and a coloring based AFCT construction method are proposed. AFCT can be constructed even the trace points change dynamically in the system runtime. Trace points can be further classified by the novel coloring algorithm, so that only a few special functions are needed to record additional tracking data, which greatly reduces the total amount of tracking data.

3) \textbf{Fault diagnosis approach}. A Graph Convolutional Network (GCN) and fault injection based fault diagnosis method is proposed. To populate the faults and their error characteristics database, we derived a comprehensive fault model at software functional level which spanning across several types of faults, including faults occured in control flow, system resource, input and output interface. Experiments on Redis, Nginx, Httpd and SQLite show that our method is faster, lower overhead and more accurate than the existing three representative fault diagnosis methods.

\section{Overview}
\begin{figure*}[tb]
\centering
\includegraphics[width=6in]{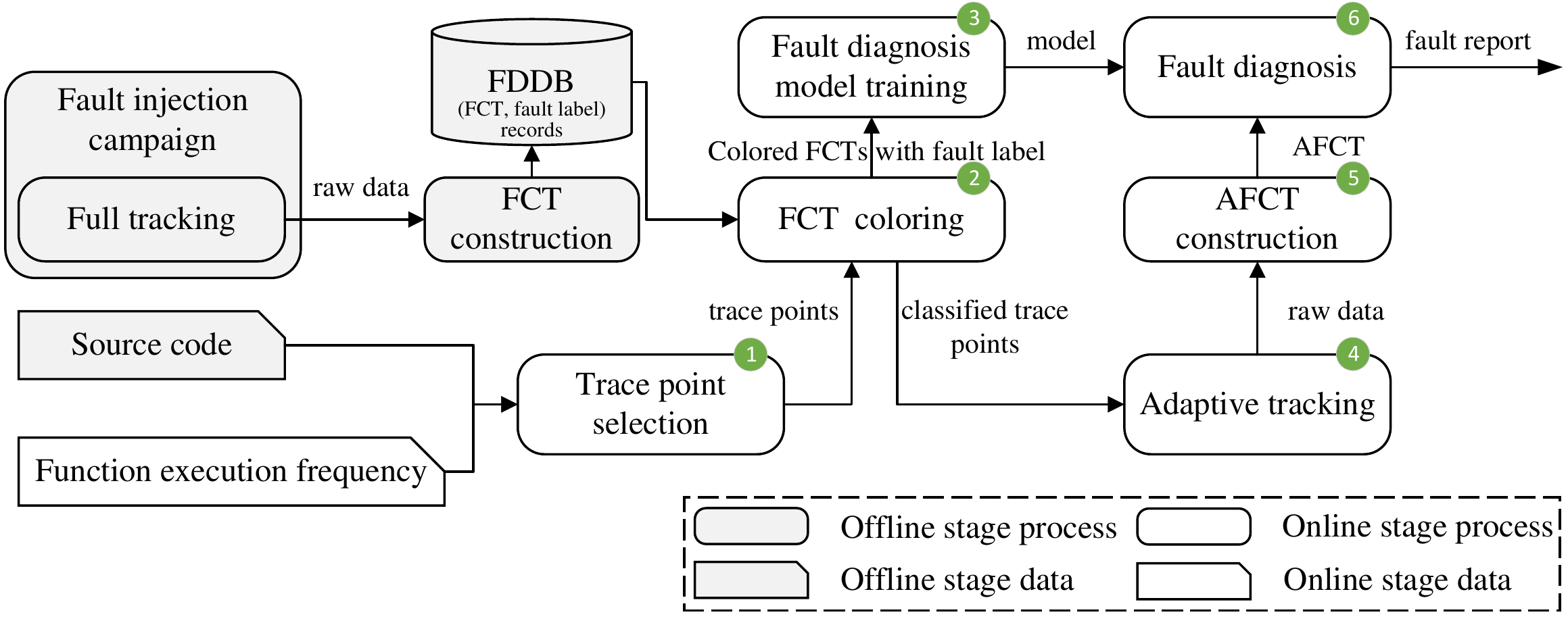}
\caption{Overview of fault diagnosis framework using AFETM.}
\label{fig_overview}
\end{figure*}

Overview of the process for fault diagnosis framework using AFETM is shown as Fig.~\ref{fig_overview}. The content can be divided into the following offline and online parts. 

In the offline part, we design and carry out fault injection campaign to inject faults into all functions of the diagnosed object to obtain its error characteristics (FCTs with fault) caused by function level faults. These FCTs with injected fault labels construct the Fault Diagnosis Database (FDDB).

The online part consists of six sub-parts. The trace points selection sub-part generates preliminary trace points according to the analysis of program source code and the function execution frequency sampling data, it determines which functions need to be tracked. In the FCT coloring sub-part, the FCTs in FDDB are colored according to the preliminary trace points. The trace points are further classified to facilitate the subsequent use of different tracking strategies to reduce the amount of trace data. The fault diagnosis model training sub-part learns AFETM fault diagnosis model from the generated colored FCTs with fault labels. Then the adaptive tracking, AFCT construction, and fault diagnosis sub-parts collect, process, and analyze the trace data respectively to report whether there are errors, where the faults are located.

The methods and related algorithms used in the above processes are described in detail in next section.

\section{Method}
\subsection{Trace Points Adaptive Selection}
\begin{figure}[tb]
\centering
\includegraphics[width=3.5in]{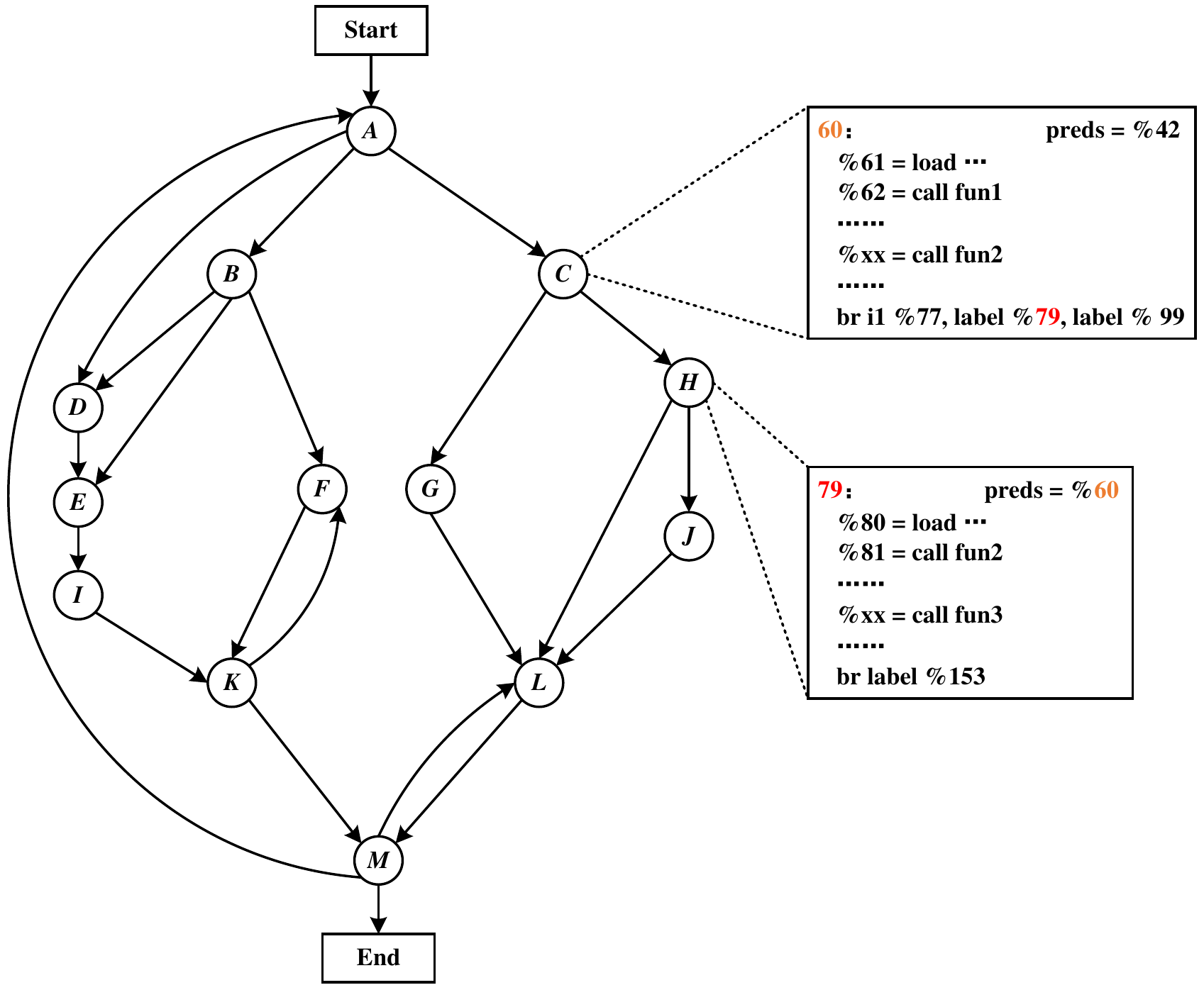}
\caption{Control flow graph generated by IR.}
\label{IR}
\end{figure}

Trace points should be selected so that the program structure coverage is as high as possible, and the overhead introduced should be controlled within an acceptable range. We analyze the function coverage at IR layer to characterize the tracking coverage of a given program. The control flow graph can be generated by parsing IR codes, as shown as Fig.~\ref{IR}. We can use common compilers, e.g., clang/LLVM, to obtain IR code corresponding to the program source code, and build its corresponding control flow graphs. Each function defined in the program can be represented by a control flow graph. Where, each node represents an IR basic block and the directed edges represent the execution order of the IR basic blocks. Fig.~\ref{IR} shows the control flow graph of a given function, in which there are thirteen nodes (node A--M). Each IR basic block (node) contains a label , e.g., node \textit{H} with label 79, and its predecessors, e.g., node \textit{C} with label 60 is predecessor of node \textit{H}. And for the instructions in each IR basic block, we concentrate on the function call instructions, which represent some functions called in the IR basic block, e.g., \textit{fun2} is called in node \textit{C} and node \textit{H}. If \textit{fun2} is traced, then the node \textit{C} and node \textit{H} are covered. So we convert the program structure coverage problem into the following Set Cover Problem (SCP) \cite{SCP}. 

Given IR basic block set \(\mathbb{B}=\{b_1,b_2,...,b_n\}\), function set \(\mathbb{F}=\{f_1,f_2,...,f_m\}\). Each \(f_j\in\mathbb{F}\) ($j=1,2,...,m$) covers a subset of \(\mathbb{B}\). The optimization goal of the SCP is to select a subset \(\mathcal{F}\subset \mathbb{F}\), which covers all basic blocks in \(\mathbb{B}\) with minimal number of functions in \(\mathbb{F}\).

In order to limit the tracking overhead within an acceptable range, we introduce weight factor into above SCP. To describe the cost of tracking \(f_j\in\mathbb{F}\), weight set \(\mathbb{W}=\{w_1,w_2,...,w_m\}\) is defined, where \(w_j\) is the cost of tracking \(f_j\). The number of executions of \(f_j\) per second is selected to represent its tracking cost, i.e., the value of \(w_j\), because tracking overhead is linearly related to function execution frequency. 

Finally, the trace points selection problem can be expressed as a Minimum Weighted Set Coverage Problem \cite{MWSCP} with Upper Bound total weight Constraint (MWSCP-UBC). The optimization goal of the MWSCP-UBC is to select a subset, \(\mathcal{F}\subset \mathbb{F}\), which covers maximum number of IR basic blocks and the total weight not exceeds the given upper bound \(w_{ub}\). Assuming \(h_{ij}\) represents whether \(b_i\) ($i=1,2,...,n, b_i\in \mathbb{B}$) covered by \(f_j\), \(h_{ij}=1\) represents \(f_j\) covers \(b_i\). \(x_j\) represents whether \(f_j\) is in solution \(\mathcal{F}\), \(x_j=1\) represents \(f_j\in \mathcal{F}\). Then the number of times the \(b_i\) was covered can be expressed as \eqref{c_i}.

\begin{equation} 
c_i=\sum_{j=1}^{m}h_{ij}x_{j}
\label{c_i}
\end{equation}
Whether \(b_i\) is coverd by solution \(\mathcal{F}\) can by expressed by \eqref{a_i}.
\begin{equation} 
a_i=\left\{
	\begin{array}{lr}
	1, & c_i\ge 1 \\
    0, & otherwise
	\end{array}
\right.
\label{a_i}
\end{equation}
The mathematical programming form of MWSCP-UBC can be expressed as follows.

\textbf{Target:}
\begin{equation} 
	\min \quad F(x)=\delta_1(1-\frac{1}{n}\sum_{i=1}^{n}a_i)+\delta_2\frac{\sum_{j=1}^{m}x_{j}w_{j}}{w_{ub}}
	\label{MWSCP_UBC_target}
\end{equation}

\textbf{Constraint condition:}
\begin{equation} 
    \sum_{j=1}^{m}x_{j}w_{j}\leq w_{ub}
    \label{MWSCP_UBC_contraint}
\end{equation}
In \eqref{MWSCP_UBC_target}, \(\delta_1=n/(n+1)\) and \(\delta_2=1/(n+1)\). These two parameters ensure that the structural coverage of tracking is the dominant factor (\(\delta_1\)) which affecting the value of fitness function, and the total cost of tracking is the secondary factor (\(\delta_2\)). Because when the \(w_{ub}\) allows full covering of all the IR basic blocks, the optimization target is to achieve minimum total weight to cover all IR basic blocks. When the \(w_{ub}\) not allows full covering of all the IR basic blocks, maximum structural coverage is the optimization target. The constraint condition requires the total weight of tracking functions in \(\mathcal{F}\) not exceeds the given upper bound \(w_{ub}\).

Same as SCP \cite{SCP}, MWSCP-UBC is an NP-hard problem. We propose a heuristic ant colony algorithm based on the Max-Min Ant System (MMAS) proposed by St\"utzle and Hoos \cite{MMAS} to solve this problem. We choose it because it is one of the top performing methods for this type of problem \cite{Ant_colony_optimization}. Compared with other traditional methods, it can search around the high quality solutions, and avoid premature convergence and falling into local optimization.

We assume that the number of ants is \(N_{ant}\) and the number of iterations is \(N_{run}\). In iteration \(t\), the pheromone on function \(f_j\) is \(\tau_{j}(t)\), the probability that ant \(h\) selects function \(f_j\) is \(P_{j}^{h}(t)\), the heuristic function is \(\eta_{j}(t)\), the pheromone heuristic factor is \(\alpha\), the expected heuristic factor is \(\beta\), the tabu list of ant \(h\) is \(tabu_{h}(t)\) (the set of functions that has been selected), the pheromone evaporation coefficient \(\rho\in(0,1)\), the pheromone increment of ant \(h\) on function \(f_j\) is \(\Delta\tau_{j}^{best}\), the set of functions that has not been selected and can be selected by ant \(h\) is \(allowed_{h}\). The transition probability function \(P_{j}^{h}(t)\) can be expressed as \eqref{MMAS_P}.

\begin{equation} 
    P_{j}^{h}(t)=\left\{
	\begin{array}{lr}
	\displaystyle\frac{\tau_{j}(t)^{\alpha}
    \eta_{j}(t)^{\beta}}{\sum_{f_j\in allowed_h}(\tau_{j}(t)^{\alpha} \eta_{j}(t)^{\beta})}, & f_j\in allowed_h\\
    0, & otherwise
	\end{array}
    \right.
    \label{MMAS_P}
\end{equation}
Where \(\alpha\) controls the importance of pheromone accumulated by ants, \(\beta\) controls the the importance of heuristic information. In iteration \(t+1\), the pheromone function of function \(f_j\) is expressed as \eqref{MMAS_tau}.

\begin{equation} 
    \tau_{j}(t+1)=(1-\rho) \tau_{j}(t)+\Delta\tau_{j}^{best}
    \label{MMAS_tau}
\end{equation}

In the first iteration, \(\tau_{j}(0)=\tau_{ini}\), \(\tau_{ini}\) is a constant value. \(1-\rho\) is the pheromone residue coefficient. In MMAS, we need to set upper and lower bounds for pheromone. We assume that \(\tau_{j}(t)\in [\tau_{min},\tau_{max}]\), the initial value of \(\tau_{max}\) is \(\tau_{j}(0)\), \(\tau_{min}=\tau_{max}/l\), \(l\) is the lower bound factor, and \(l>1\). The pheromone increment \(\Delta\tau_{j}^{best}\) on function \(f_j\) is expressed as \eqref{MMAS_delta}.

\begin{equation} 
    \Delta\tau_{j}^{best}(t)=\left\{
	\begin{array}{lr}
	\displaystyle\frac{Q}{N_{ant}F(x^{best})\tau_{j}(t)}, & x_{j}^{best} = 1\\
    0, & otherwise
	\end{array}
    \right.
    \label{MMAS_delta}
\end{equation}
where \(x^{best}\) is the optimal solution vector of this model. The value of \(F(x^{best})\) can be calculated by \eqref{MWSCP_UBC_target}. In traditional MMAS, pheromone increment equals to the reciprocal of fitness function value in the optimal solution. However, in our algorithm, it is proportional to to the reciprocal of fitness function value in the optimal solution divided by the pheromone function. It can effectively prevent the algorithm from falling into local optimization prematurely. The total number of pheromone is \(Q\). \(\eta_{j}(t)\) is the heuristic function, it represents the expected degree of selecting function \(f_j\), and expressed as \eqref{MMAS_eta}.

\begin{equation} 
    \eta_{j}(t)=\left\{
	\begin{array}{lr}
	\displaystyle\frac{1}{w_j}, & w_j\ne 0\\
    \eta^{max}, & otherwise
	\end{array}
    \right.
    \label{MMAS_eta}
\end{equation}
Where \(w_j\) is the cost of tracking function \(f_j\), and it is non-negative. If \(w_j=0\) then \(\eta_{j}(t)=\eta^{ max}\) (\(>1\)). It indicates that ant prefers functions with lower cost.

The MMAS based algorithm for trace points selection is shown as Algorithm~\ref{MMAS_algorithm}. Its input contains $\mathbb{B}$, $\mathbb{F}$, $\mathbb{W}$, $w_{ub}$, $N_{ant}$, $N_{run}$, $\alpha$, $\beta$, $\rho$, $\tau_{ini}$, $\eta^{max}$, $Q$, $l$, and output contains $x^{best}$, $F(x^{best})$.

\begin{figure}[tb]
\removelatexerror
\begin{algorithm}[H]
    \label{MMAS_algorithm}
    \caption{MMAS based trace points selection}
    \LinesNumbered
    \KwIn {$\mathbb{B}, \mathbb{F}, \mathbb{W}, w_{ub}, N_{ant}, N_{run}, \alpha, \beta, \rho, \tau_{ini}, \eta^{max}, Q, l$}
    \KwOut {$x^{best}, F(x^{best})$}
    $x^{best} \leftarrow 0$\;
    $\tau \leftarrow \tau_{ini}$\;
    \For{$t=1$ \KwTo $N_{run}$} {
        $x^{ib} \leftarrow 0$\;
        \For{$h=1$ \KwTo $N_{ant}$} {
            $allowed \leftarrow 1$\;
            $w_{total} \leftarrow 0$\;
            \While{$w_{total} < w_{ub}$} {
                select funtion $f_j$ from $\mathbb{F}$ by roulette wheel selection algorithm according to \eqref{MMAS_P}\;
                $w_{total} \leftarrow w_{total}+w_j$\;
                \If{$w_{total} \leq w_{ub}$} {
                    $allowed[j] \leftarrow 0$\;
                }
            }
            $x \leftarrow 1-allowed$\;
            \If{$F(x)<F(x^{ib})$} {
                $x^{ib} \leftarrow x$\;
            }
        }
        update $\Delta\tau^{best}(t), \tau(t)$ according to \eqref{MMAS_tau} and \eqref{MMAS_delta}\;
        \uIf{$\tau(t) < \tau_{min}$} {
            $\tau(t) \leftarrow \tau_{min}$\;
        } \ElseIf{$\tau(t) > \tau_{max}$} {
            $\tau(t) \leftarrow \tau_{max}$\;
        }
        \If{$F(x^{ib})<F(x^{best})$} {
            $x^{best} \leftarrow x^{ib}$\;
        }
    }
    calculate $F(x^{best})$ according to \eqref{MWSCP_UBC_target}\;
    \Return {$x^{best}, F(x^{best})$}\;
\end{algorithm}
\end{figure}

This algorithm can be divided into the following steps.

\textbf{Step 1}. Initialize parameters. The optimal trace points selection vector $x^{best}$ is initialized to zero vector, the pheromone vector $\tau$ is initialized to $\tau_{ini}$. (line 1-2)

\textbf{Step 2}. Each ant selects functions from $allowed$ by roulette wheel selection algorithm according to \eqref{MMAS_P} until the total weight $w_{total}$ reaches the upper bound $w_{ub}$. (line 6-12)

\textbf{Step 3}. In each iteration of $N_{run}$, the iteration best solution $x^{ib}$ is computed (line 4-15). Then, $\Delta\tau^{best}(t)$ and $\tau(t)$ are updated according to \eqref{MMAS_tau} and \eqref{MMAS_delta} with $x^{ib}$, and $\tau(t)$ is limited within $[\tau_{min},\tau_{max}]$ (line 17-20). If the fitness value of iteration best solution smaller than that of global best solution, then update $x^{best}=x^{ib}$ (line 21-22).

\textbf{Step 4}. When the iteration number $t$ reaches $N_{run}$, calculate $F(x^{best})$ according to the optimal global solution $x^{best}$ and return. (line 23-24)

The time complexity of Algorithm~\ref{MMAS_algorithm} is $O(N_{run}\times N_{ant}\times m)$, and the space complexity is $O(m)$, where $m$ is the size of $\mathbb{F}$. 

\subsection{AFCT Construction}

Comparing with function call sequence, FCT keeps more structural characteristics of program. However, the traditional FCT construction methods require all functions defined in the program to be tracked, they are not suitable for adaptive tracking. If all functions are traced, FCT can be constructed according to the \textit{caller} attribute. Otherwise, if not all functions are traced, function call forest is obtained, which leads to the loss of structural information. 

To solve the above problem, we propose a method to construct the AFCT, which can be used when tracking a subset of all functions. The key is to record \textit{callstack} attribute, which can be used to help us complete a part of missing functions and their call relations. But it is not advisable to record \textit{callstack} attribute of all functions because of its large size. Therefore, the first step is to select a subset from the trace point set to let them record the \textit{callstack} attribute. The second step is to construct AFCT from the collected trace data.

\subsubsection{FCT Coloring} In the testing phase, we can obtain the FCT set for the normal operation of the software. Based on this FCT set and the obtained trace point set, we propose a coloring algorithm to classify these trace points. It can be used to determine which functions should record the \textit{callstack} attribute. 

The FCT coloring algorithm is shown as Algorithm~\ref{FCT_COLORING}. Its input contains $FCTs$, $\mathcal{F}$, and output contains $FCTs^{colored}$, $\mathcal{F}^{callstack}$. $FCTs$ is the FCT set for the normal operation of the software. $\mathcal{F}$ is the trace point set obtained by Algorithm~\ref{MMAS_algorithm}. $FCTs^{colored}$ is colored FCT set, which color attributes of each node in FCT are filled with ``white'', ``red'' or ``blue''. The color attribute is used to obtain $\mathcal{F}^{callstack}$ and to help the following fault diagnosis process. $\mathcal{F}^{callstack}$ is the function set selected from $\mathcal{F}$, and the \textit{callstack} attribute of each function in $\mathcal{F}^{callstack}$ should be recorded. 

This algorithm can be divided into the following steps.

\textbf{Step 1}. Initialize parameters. The \textit{callstack} recording function set $\mathcal{F}^{callstack}$ is initialized to empty. (line 1)

\textbf{Step 2}. Initialize color attribute of each node. For each FCT, traverse it in preorder sequence, if the function represented by node is in trace point set $\mathcal{F}$ then mark it to ``white'', otherwise mark it to ``red''. (line 2-6)

\textbf{Step 3}. Adjust color attribute for "red" nodes. Traverse each node same as Setp 2, for each red node, if it descendent exist "white" ones, then change the ``red'' node to ``blue'', and add its nearest ``white'' descendent to $\mathcal{F}^{callstack}$. (line 7-11)

The time complexity of Algorithm~\ref{FCT_COLORING} is $O(N_{node})$, $N_{node}$ is the total number of nodes in all $FCTs$, and the space complexity is $O(m)$, where $m$ is the size of $\mathbb{F}$. 

\begin{figure}[tb]
\removelatexerror
\begin{algorithm}[H]
    \label{FCT_COLORING}
    \caption{FCT coloring}
    \LinesNumbered
    \KwIn {$FCTs$, $\mathcal{F}$}
    \KwOut {$FCTs^{colored}$, $\mathcal{F}^{callstack}$}
    $\mathcal{F}^{callstack} \leftarrow \varnothing$\;
    \For{$FCT$ \textbf{in} $FCTs$} {
        \For{$node$ \textbf{in} $FCT.iter()$}  { 
            \If{$node$ \textbf{in} $\mathcal{F}$} {
                $node.color \leftarrow$ white\;
            }
            \Else {
                $node.color \leftarrow$ red\;
            }
        }
        \For{$node$ \textbf{in} $FCT.iter()$} {
            \If{$node.color$ == red} {
                \For{$node\_tmp$ \textbf{in} $node.iter()$} {
                    \If{$node\_tmp.color$ == white} {
                        $node.color \leftarrow$ blue\;
                        $\mathcal{F}^{callstack}.add(node\_tmp)$\;
                        \textbf{break}\
                    }
                }   
            }
        }
    }
    \Return {$FCTs^{colored}$, $\mathcal{F}^{callstack}$}\;
\end{algorithm}
\end{figure}

\subsubsection{AFCT Construction}

The AFCT construction algorithm is shown as Algorithm~\ref{AFCT_construction}. Its input is $trace\_data$, and output is the AFCT set $AFCTs$. Where $trace\_data$ is the collected raw data while using adaptive tracking, which is sequence data composed of $(function, caller, callstack)$ elements in time serial.

This algorithm can be divided into the following steps.

\textbf{Step 1}. Initialize parameters. The AFCT set $AFCTs$ is initialized to empty set, the temporary set $new\_nodes$ used to store new added tree node is also initialized to empty set. (line 1-2)

\textbf{Step 2}. Find root nodes from the $trace\_data$ and add them into set $AFCTs$. If a function does not exist caller, then it is a root node. (line 3-8)

\textbf{Step 3}. Construct function call forest from the $trace\_data$. Put the root nodes into set $new\_nodes$ and find their children nodes. Add these children nodes into $new\_nodes$ and remove their parent nodes until the set $new\_nodes$ becomes empty. Function $not\_pred(new\_node,function)$ return true if $function$ is an ancestor of $new\_node$, it is used to prevent dead loops when handling recursive calls. (line 9-16)

\textbf{Step 4}. Merge sub-trees to get AFCT. For each sub-tree, if existing a child node of its root recorded $callstrack$ attribute, then the sub-tree need to be merged. Function $merge(AFCTs,root,callstack)$ merges $root$ into $AFCTs$ according to $callstack$. (line 17-22)

The time complexity of Algorithm~\ref{AFCT_construction} is $O(N_{trace\_data}^2)$, $N_{trace\_data}$ is the length of trace data, and the space complexity is $O(N_{trace\_data})$. 

\begin{figure}[tb]
\removelatexerror
\begin{algorithm}[H]
    \label{AFCT_construction}
    \caption{AFCT construction}
    \LinesNumbered
    \KwIn {$trace\_data$}
    \KwOut {$AFCTs$}
    $AFCTs \leftarrow \varnothing$\;
    $new\_nodes \leftarrow \varnothing$\;
    \For{$function$, $caller$, $callstack$ \textbf{in} $trace\_data$} {
        \If{$caller$ \textbf{not in} $AFCTs$} {
            $AFCTs.add(caller)$\;
        }
    }
    \For{$function$, $caller$, $callstack$ \textbf{in} $trace\_data$} {
        \If{$function$ \textbf{in} $AFCTs$} {
            $AFCTs.remove(function)$\;
        }
    }
    $new\_nodes.add(AFCTs)$\;
    \While{$new\_nodes \ne \varnothing$} {
        \For{$new\_node$ \textbf{in} $new\_nodes$} {
            \For{$function$, $caller$, $callstack$ \textbf{in} $trace\_data$} {
                \If{$caller$==$new\_node$ \textbf{and} $function$ \textbf{not in} $new\_node.children$ \textbf{and} $not\_pred(new\_node, function)$} {
                     $new\_node.children.add(function)$\;
                     $new\_nodes.add(function)$\;
                }
            }
            $new\_nodes.remove(new\_node)$\;
        }
    }
    \For{$root$ \textbf{in} $AFCTs$} {
        \For{$child$ \textbf{in} $root.children$} {
            \For{$function$, $caller$, $callstack$ \textbf{in} $trace\_data$} {
                \If{$root$ == $caller$ \textbf{and} $child$ == $function$ \textbf{and} $not\_empty(callstack)$} {
                    $merge(AFCTS,root,callstack)$\;
                }
            }
        }
    }
    \Return {$AFCTs$}\;
\end{algorithm}
\end{figure}

After the AFCT construction, we can use it to do the data analysis work, i.e., fault diagnosis. Because according to the Algorithm~\ref{FCT_COLORING} and Algorithm~\ref{AFCT_construction}, all of the white nodes and blue nodes are preserved in AFCT. Although the red nodes are lost, the tree structure is still maintained. So when the red nodes in colored FCT are masked, the AFCT is equivalent to it. Therefore, AFCT can be used as an alternative to FCT.

\subsection{Fault Diagnosis}

In order to realize function level fault diagnosis, we design and carry out fault injection campaign to inject faults into all functions of the diagnosed object to obtain its error characteristics (FCTs with fault) caused by function level faults. These FCTs with injected fault labels construct the Fault Diagnosis Database (FDDB). Then, the runtime constructed AFCT can be used to do the fault diagnosis based on the colored FCTs (obtained by Algorithm~\ref{FCT_COLORING}) of FDDB. 

\subsubsection{Fault Injection}

The fault injection campaigns are designed and carried out to obtain the FDDB data. The following fault model defines three important aspects, namely, what to inject (i.e., which kind of fault), when to inject (i.e., the timing of the injection), and where to inject (i.e., the part of the system targeted by the injection) \cite{sfi_survey}. 

Specifically, five types of faults are defined in three different aspects: control, resource and interface.

\begin{itemize}
  \item Control: IP+N, process crash.
  \item Resource: deadlock.
  \item Interface: input corruption, output corruption.
\end{itemize}

They are selected because they are typical fault types with different error propagation and manifestation characteristics. IP+N and process crash usually have short error propagation and give little error information. The difference is that IP+N may causes strange control flow and fast failure, while process crash causes immediate failure. Deadlock usually leads to process/thread hangs. Interface corruption usually leads to long error propagation comparing with control type faults.

Location type of fault injection is defined at function level: the beginning of the function or the end of the function, i.e., return, because fault injection in functional granularity is usually sufficient for fault location. And for convenient, time type of faults is defined as permanent to emulate the permanent nature of software faults \cite{swf_permanent_nature}.

In practice, IP+N can be injected by modifying the value of RIP in structure CONTEXT-\textgreater uregs (for x86\_64). Process crash can be injected by raising a SIGKILL signal. Deadlock can be injected by raising a SIGSTOP signal (simulate process/thread hangs). Interface faults can be injected by flipping a bit of function input parameters or function output data. This form of fault injection is referred to as “interface error injection” or “failure injection” in some studies. Other studies refer to it as “fault injection,” since corrupted values can be seen as “external faults,” according to the definitions of Avizienis et al. \cite{sfi_survey,concept_definition}. These injection experiments can be conducted by SystemTap toolkit\cite{systemtap} with its guru mode (using -g option in stap command), which can be used to instrument the target software without stopping it.

All functions of the diagnosis target (or a set of interested functions) are the fault injection targets. And for each $(function, fault)$ pair, unique $FCTs$ (unique error characteristics) can be obtained from the fault injection campaign. Finally, the FDDB is populated with these $(function, fault, FCTs)$ data.

\subsubsection{Fault Diagnosis Model}

\begin{figure}[tb]
\centering
\includegraphics[width=3.5in]{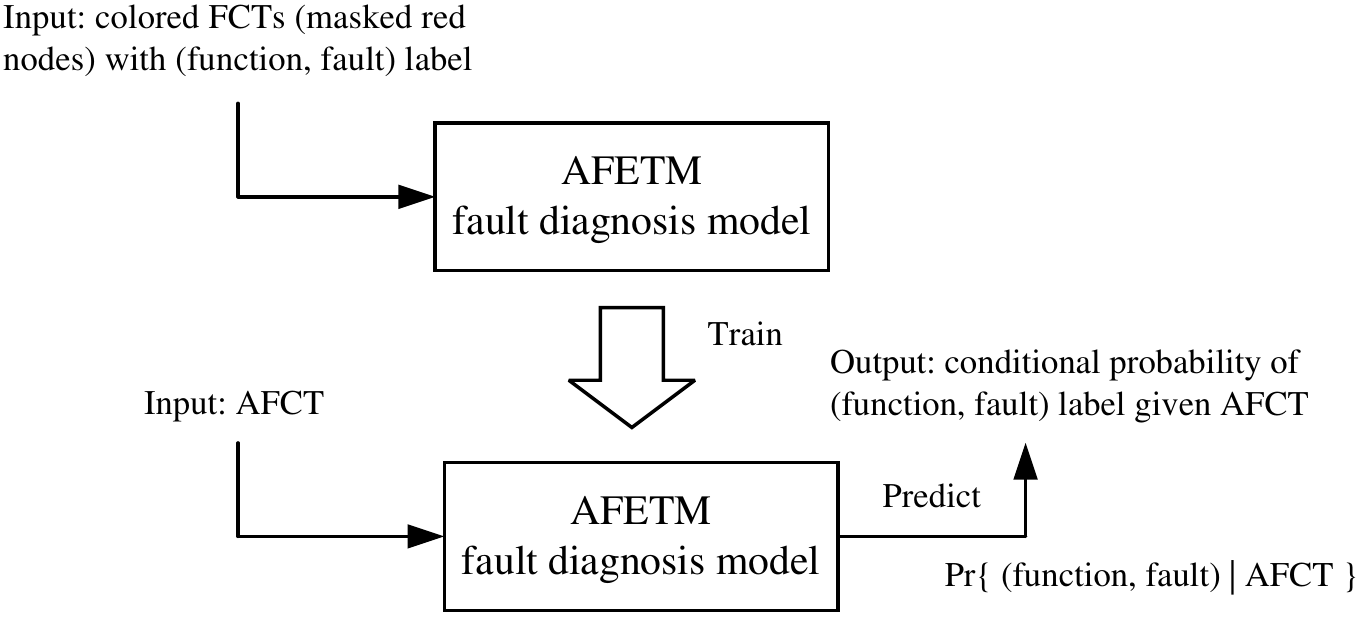}
\caption{Overview of the AFETM fault diagnosis model.}
\label{fig_fault_diagnosis_overview}
\end{figure}

The fault diagnosis problem can be converted into a classification problem: given the AFCT, how to predict the fault function or normal. We design a neural network model based on graph convolution \cite{GCN} to solve this problem. Because for the graph-level classification problem, GCN is one of the top performing method and computationally efficient \cite{GCN_review}. The overview of the AFETM fault diagnosis model is shown as Fig.~\ref{fig_fault_diagnosis_overview}. In the model training phase, the input is the colored FCTs obtained from FDDB through Algorithm~\ref{FCT_COLORING}, and we mask their red nodes because they are not exist in AFCT. In the runtime phase, the input is the constructed AFCT, and the output is the predicted conditional probability of $(function, fault)$.

The layer-wise propagation rule of the graph convolutional operation is shown as \eqref{GCN_transition_fun}. The $\Tilde{A}=A+I$ is the adjacency matrix of the tree (here, it is an AFCT or colored FCT) with self-connections, $I$ is the identity matrix. $\Tilde{D}_{ii}=\sum_{j=0}{\Tilde{A}_{ij}}$ and $H^{(l)}$ is a layer-specific trainable weight matrix. $\sigma(\cdot)$ is the activation function, here we choose $ReLU(\cdot)=max(0,\cdot)$. $H^{(l)}$ is the matrix of activations in the $l^{th}$ layer and $H^{(0)}$ is the input, i.e. the graph embedding of AFCT or colored FCT.

\begin{equation} 
    H^{(l+1)}=\sigma(\Tilde{D}^{-\frac{1}{2}} \Tilde{A} \Tilde{D}^{-\frac{1}{2}} H^{(l)}W^{(l)})
    \label{GCN_transition_fun}
\end{equation}

The node-wise propagation rule of the graph convolutional operation is shown as \eqref{GCN_transition_fun_node}. where, $\Tilde{d}_i=1+\sum_{j\in\mathcal{N}(i)}e_{j,i}$, $e_{j,i}$ is the edge weight from source node $j$ to target node $i$ (we use default value 1.0), $\mathcal{N}(\cdot)$ is the neighbor set of the node. Each convolution layer makes all nodes integrate the features of their neighbor nodes and edges. It is generally considered that the graph convolution operation does not need too many layers, and three layers are usually enough.

\begin{equation} 
    H^{(l+1)}_{i}=W^{(l)\top} \sum_{j\in\mathcal{N}(v)\cup\{i\}} \frac{e_{j,i}}{\sqrt{\Tilde{d}_j\Tilde{d}_i}}H^{(l)}_j
    \label{GCN_transition_fun_node}
\end{equation}

After the softmax operation, the output can be obtained, it is a vector which contains the probability of each classification type, i.e., the probability of $(function, fault)s$ and normal. The highest probability or TopK results can be used as the final diagnosis result.

\section{Evaluation}

To evaluate AFETM, we consider the following research questions:

\textbf{RQ1:} How effective is AFETM at diagnosing software fault? 

\textbf{RQ2:} How does it affect the performance of the target system? 

\textbf{RQ3:} How does it compare with existing methods?

\subsection{Environment}
\textbf{Target system:} In order to evaluate AFETM comprehensively, we select Redis, Nginx, Httpd, SQlite as the experiment targets to demonstrate the benefits of adaptive tracking for fault diagnosis. We choose above four software because they are the typical representatives which are widely used in web service area. Among them, Redis is a high-performance data cache software. We chose it to verify the effectiveness of the adaptive tracking for very high-performance software. Httpd and Nginx are the most famous open-source HTTP software and load balancing software, respectively. SQLite is the most used database engine in the world.

\begin{figure}[tb]
\centering
\includegraphics[width=3.3in]{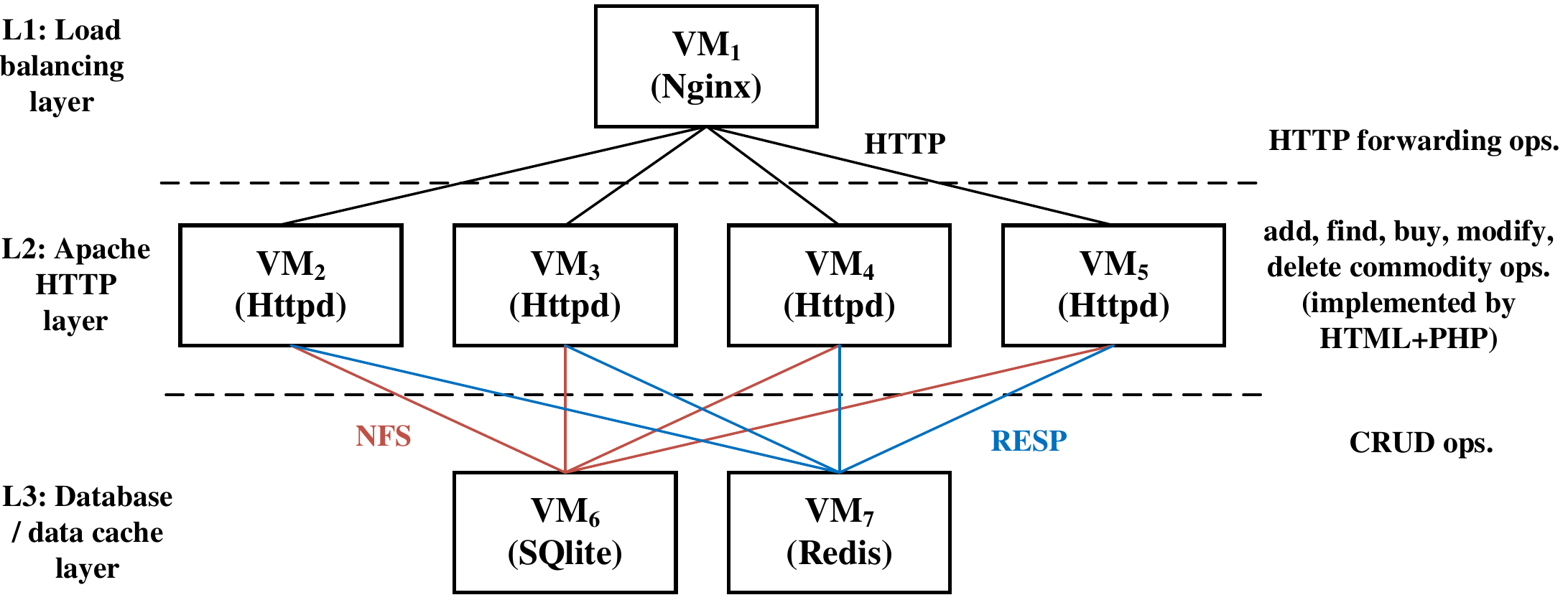}
\caption{Three-layer online shop web structure based on Nginx, Httpd, Redis and SQlite.}
\label{fig_4}
\end{figure}

\textbf{Benchmark:} The Nginx, Httpd, SQlite and Redis based Online Shop Web (NHSR-OSW) benchmark similar to TPC-W \cite{TPCW} and RUBiS \cite{RUBIS} is constructed. As shown as Fig.~\ref{fig_4}, NHSR-OSW is deployed on seven virtual machines in three layers. The physical machine configuration is Intel Core i7-8700 Processor (3.20GHz), 16GB memory, and 1TB hard disk (7200RPM). The virtual machine (VM) configuration is single core and single thread processor, 1GB memory, 100GB hard disk. In the load balancing layer (L1), a Nginx service is deployed on VM$_{1}$ to complete the HTTP requests forwarding operations. In the Apache HTTP layer (L2), four Httpd services are deployed on VM$_{2}$-VM$_{5}$ to complete the add, find, buy, modify and delete operations for commodity. The commodity management program are written in HTML and PHP. In the database/data cache layer (L3), a SQlite service and a Redis service are deployed on VM$_{6}$ and VM$_{7}$ respectively to complete the create, retrieve, update and delete (CRUD) operations for data record. NFS protocol is used to mount the directory where the database file is located to VM$_{2}$-VM$_{5}$, because SQlite does not natively support remote access. Httpd service communicate with Redis service by phpredis (version 5.3.4) module based on Redis serialization protocol (RESP).

\textbf{Fault Injection Campaign:} To populate FDDB for NHSR-OSW, we need to design and carry out fault injection campaigns. The fault model has been presented in above section. The fault injection number of IP+N, process crash, and deadlock are same as the number of distinct functions which are activated under the workload, i.e., NHSR-OSW benchmark. The fault injection number of interface input corruption and interface output corruption are same as the number of function parameters and return data, respectively. The Httpd functions have 4 parameters in average while the Redis functions have only 2 parameters in average. Therefore, the number of fault injections of interface input corruption for Httpd is approximately twice than Redis. For NHSR-OSW, 6696 faults were constructed. For each fault, we inject it several times, until there is no new FCT generated.

\subsection{Metrics and Comparison Targets}
\textbf{Metrics:} The following metrics are used to analyze the fault diagnosis capability and overhead of AFETM.

\begin{itemize}
    \item The Error Detection Rate (EDR) represents the probability that the error is perceived. It is expressed as \eqref{EDR}.
    \begin{equation} 
        EDR=\frac{N_{error}}{N_{injection}} \times 100\%
        \label{EDR}
    \end{equation}
    \item The Fault Location Rate (FLR) represents the probability that the fault is located. It is expressed as \eqref{FLR}.
    \begin{equation} 
        FLR=\frac{N_{location}}{N_{injection}} \times 100\%
        \label{FLR} 
    \end{equation}
    Where, $N_{injection}$ is the number of fault injection experiments, and in each experiment one fault is injected. $N_{error}$ is the number of experiments in which error was found by the fault diagnosis method. $N_{location}$ is the number of experiments in which root fault was located by the fault diagnosis method.
    \item The Fault Location Distance (FLD) represents the distance between the located fault and the real fault. It is calculated by traversing FCT in preoder sequence. 
    \item The Fault Diagnosis Delay (FDD) represents the time delay from the occurrence to the report of the fault. 
    \item The Memory Occupation (MO) represents the actual memory size occupied by the fault diagnosis method.
    \item The Response Time Growth Rate (RTGR) represents the percentage increase in response time of the target system caused by the fault diagnosis method. It is expressed as \eqref{RTGR}.
    \begin{equation} 
        RTGR=\frac{T_{after}-T_{before}}{T_{before}}\times 100\% 
        \label{RTGR}
    \end{equation}
    Where $T_{before}$ is the response time of the target system before using the fault diagnosis method, and $T_{after}$ is the response time of the target system using the fault diagnosis method.
\end{itemize}

\textbf{Comparison targets:} Three representative methods are selected to compare with the method proposed in this paper.

\begin{itemize}
  \item Log based method: There are too many log based fault diagnosis methods. We use an idealized method to represent these kind of methods. If error log messages are found after a fault injection, then the injected fault is marked as covered, and we assume its FLR is equal to EDR. Take Reids for example, anetSetError(), serverLog(), RM\_LogRaw(), rdbCheckThenExit(), rdbCheckError(), or rdbCheckSetError() will be called to log the error information when Redis recognize an error state. For serverLog() and RM\_LogRaw(), the log level parameter can be used to distinguish error log. The distance from err\_log() function to the fault injected function can be calculated. Generally, the bigger the distance, the more difficult it is to locate the fault, so it is used to represent the FLD. 
  
  \item FCT-EDC method \cite{trace_overhead}: Wang et al. collect the FCT\_run by full tracking and calculate its tree edit distance with FCT\_normal. If the distance exceeds threshold, it is considered that the system is in error state. And the fault is located by comparing the preorder traversal sequence of FCT\_run and FCT\_normal.
  
  \item AFCT-GDC method: Pham et al. perform component level fault diagnosis based on (fault, FCTs) database \cite{target_fault_injection}. But it cannot be directly used for comparison with our method. We combine its Gaussian influence with our AFCT data model to enable it to perform function level fault diagnosis. The Gaussian influence is computed as \eqref{Gaussian_influence} \cite{target_fault_injection}. Where, $t_r$ is the AFCT, $t_{db}$ is FCTs of (fault, FCTs), and $\sigma$ is the standard deviation of the pair-wise distances of all (fault, FCTs).
  
  \begin{equation} 
    \Delta_{Gaussian}(t_r,t_{db})=e^{-\displaystyle\frac{\delta(t_r,t_{db})^2}{2\sigma^2}}
    \label{Gaussian_influence}
  \end{equation}
  
\end{itemize}

\subsection{Experimental Results and Analysis}

\begin{figure*}[tb]
\centering
\includegraphics[width=6.0in]{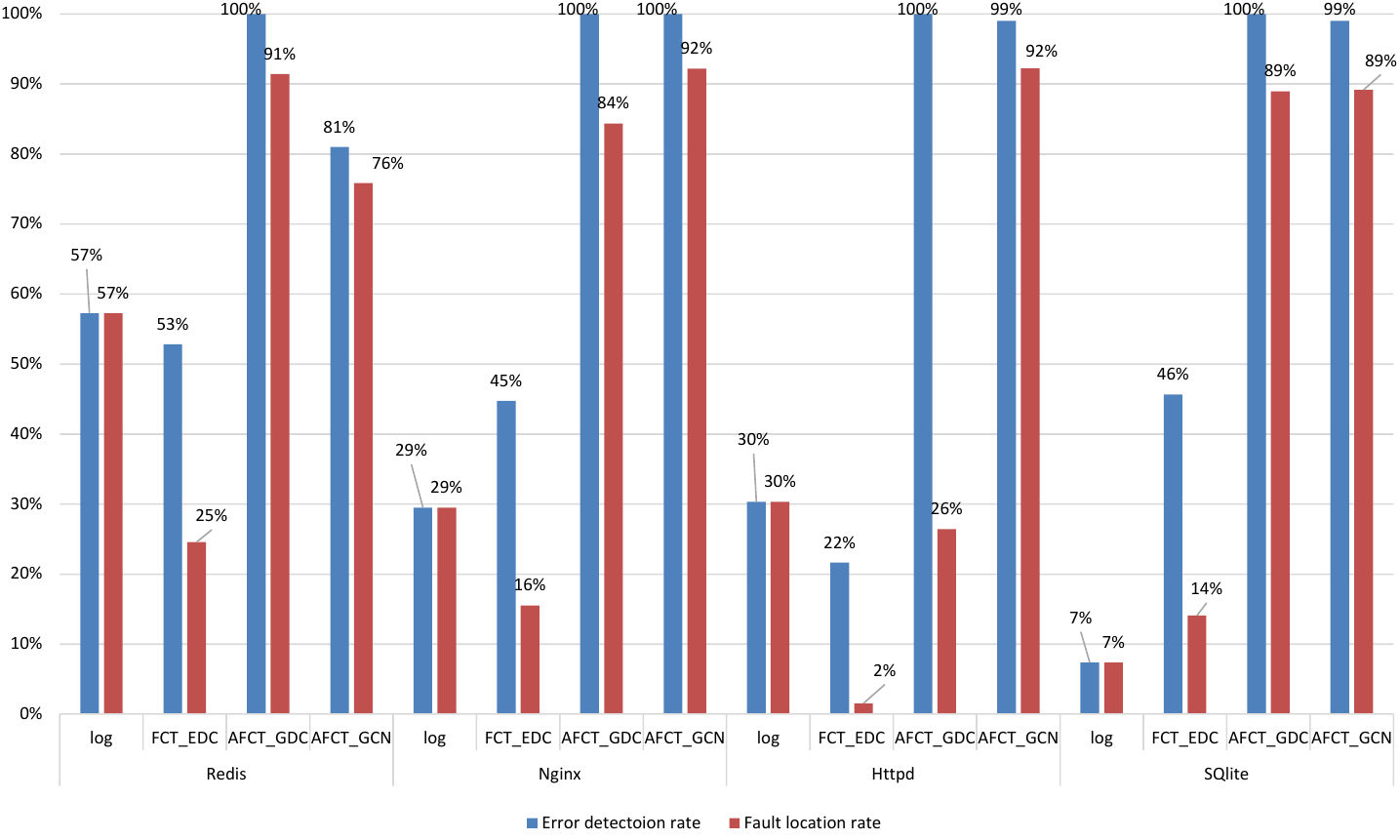}
\caption{Error detection rate and fault location rate for NHSR-OSR using different fault diagnosis methods.}
\label{edr_flr}
\end{figure*}

\textbf{RQ1 \& RQ3: Effectiveness and comparison.} The EDR and FLR for NHSR-OSR using different fault diagnosis methods are shown as Fig.~\ref{edr_flr}. From the four groups of experiments on Redis, Nginx, Httpd and SQLite, it can be seen that the fault diagnosis method with the highest EDR is AFCT\_GDC which combines the AFCT data model proposed in this paper and Gaussian inference \cite{target_fault_injection}. The EDR of this method reached 100\% in four groups of experiments. The second highest EDR is obtained by our AFETM (AFCT\_GCN) method. It achieves 81\% on Redis, 100\% on Nginx, 99\% on Httpd and SQLite. The above experimental results indicate that the AFCT data model and its construction method proposed in this paper can maintain the error feature information sufficiently. The EDR of log based method and FCT\_EDC method are not ideal in this fault injection campaign, and their EDRs on Nginx, Httpd and SQLite are even less than 50\%.

The AFETM method proposed in this paper has the highest FLR on Nginx, Httpd, and SQlite, which value are 92\%, 92\%, and 89\%, respectively. The AFCT\_GDC method has the highest FLR (91\%) on Redis. But it is only 26\% on Httpd. The above experimental data show that the methods based on Gaussian inference and GCN have higher fault location ability. Especially, the GCN method proposed in this paper has the smallest difference between FLR and EDR except for the log based method (it assumes that FLR is equal to EDR). It means that the most of errors (92\% in average) that can be perceived which fault can be located. The FLR of FCT\_EDC method is the smallest on Redis (25\%), Nginx (16\%) and Httpd (2\%).

\begin{figure}[tb]
\centering
\includegraphics[width=3.5in]{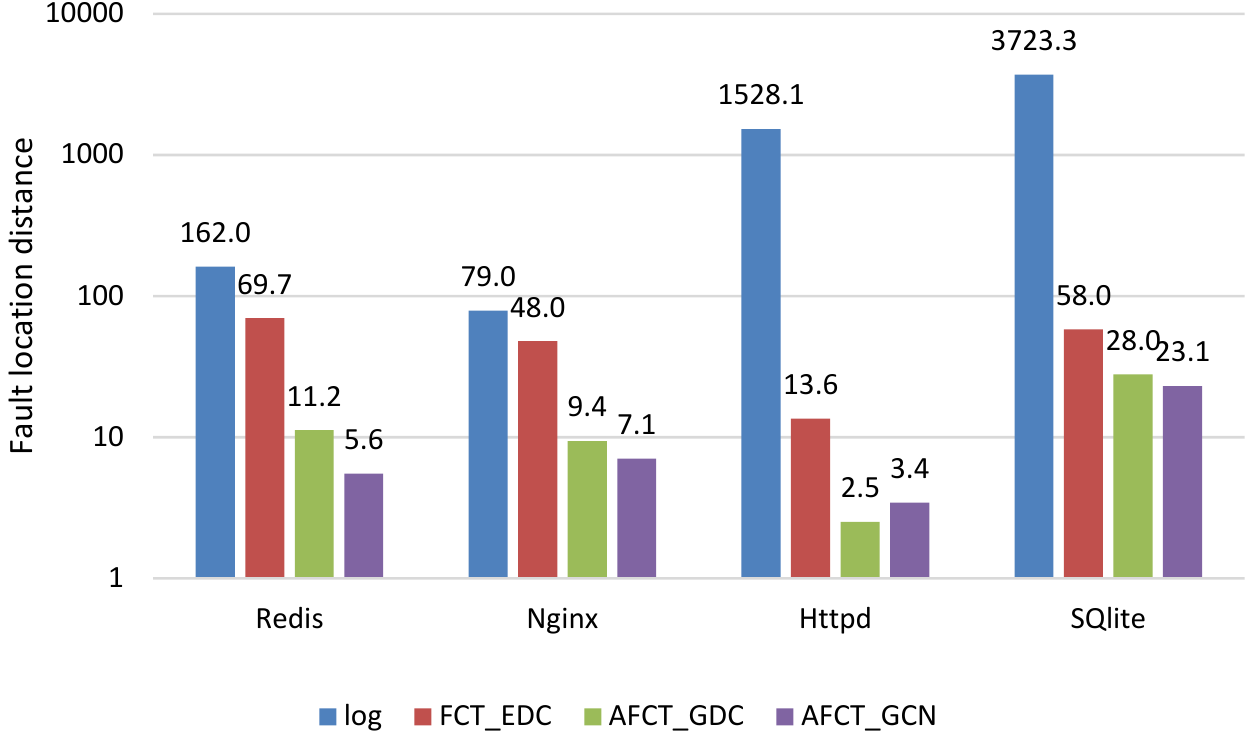}
\caption{Fault location distance for NHSR-OSR.}
\label{fld}
\end{figure}

FLD shows the accuracy of fault location. The smaller the FLD, the closer the fault location given by the fault diagnosis method is to the real fault. That means when FLD equals 0, the root cause fault location can be pinpointed. As shown as Fig.~\ref{fld}, the FLD of log based method is the largest, which means it has largest deviation in fault location. Intuitively, the bigger distance between the error log recording location and the real fault location will make it more difficult to record the real fault cause. The AFETM method obtained the minimum FLD on Redis (5.6), Nginx (7.1), and SQlite (23.1). The AFCT\_GDC method has obtained the minimum FLD on Httpd (2.5). The fault location closest to the real fault location can be obtained by these methods. FLR of AFETM on Httpd is 3.4, which means the real fault location is in around of the give fault location within 3.4 function calls in average.

\begin{figure}[tb]
\centering
\includegraphics[width=3.5in]{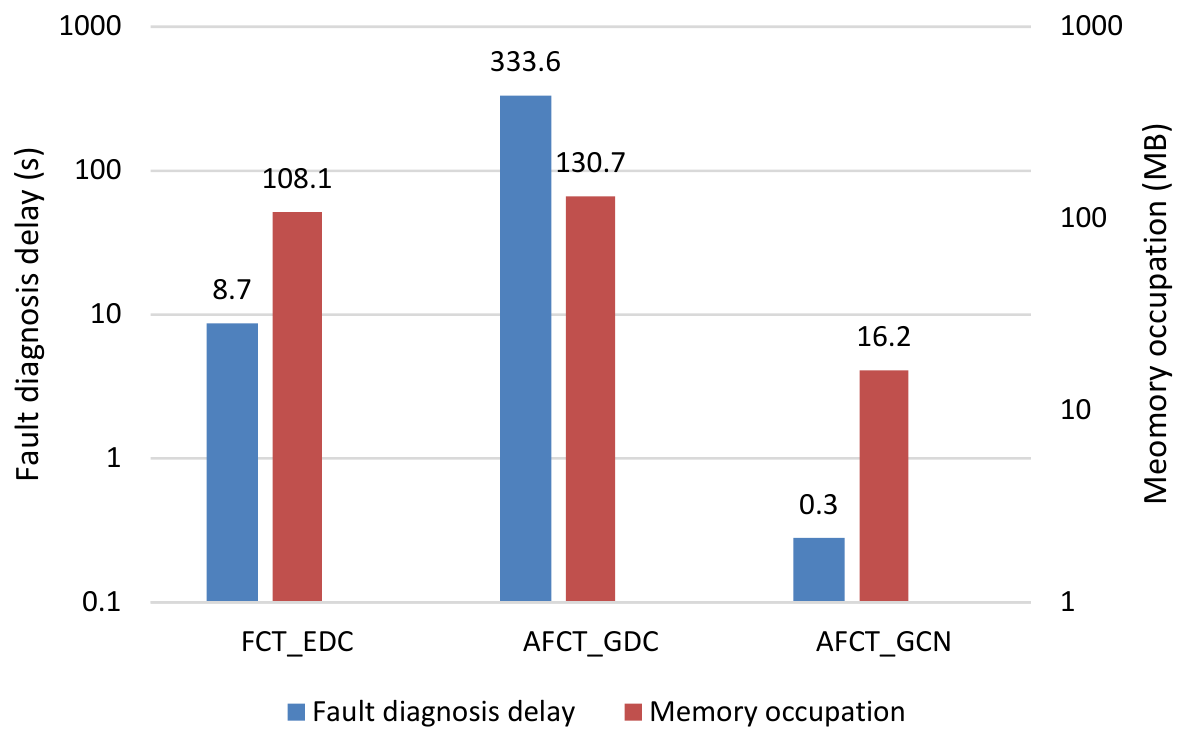}
\caption{Fault diagnosis delay and memory occupation for NHSR-OSR.}
\label{fdd_mo}
\end{figure}

FDD shows the sensitivity of the fault diagnosis system, that is, how long after the fault occurs, it can be perceived and located by the fault diagnosis system. The log based method is not compared because it is a idealized  method. As shown as Fig.~\ref{fdd_mo}, the FDD of AFETM is the smallest, with an average of 0.3 seconds, of which 0.25 seconds is used for the AFCT construction, and 0.05 seconds is used to determine whether there is a fault and progress fault location. Since the FCT\_EDC method needs to calculate the tree editing distance between the FCT generated at runtime and the normal FCTs one by one, it introduces large overhead. AFCT\_GDC has nice data on the previous EDR, FDR, FLD indicators, but its FDD is the highest. It average diagnosis time of single fault is 333.6 seconds. Because this method needs to calculate the tree editing distance between the AFCT generated at runtime and the FCTs corresponding to all faults in the fault database to obtain the Gaussian distance to judge whether there is a fault. Its time complexity is linearly increasing with the size of the fault database, i. e.,  the number of faults and the number of FCTs corresponding to each fault. The calculation of tree editing distance is a time consuming operation, so the time cost of the above two methods is much higher than that of the AFETM method.

\textbf{RQ2 \& RQ3: Overhead and comparison.} MO shows the memory cost during the operation of the fault diagnosis system. The model constructed by AFETM method is usually only about 10$^+$MB in size. As shown as Fig.~\ref{fdd_mo}, in the actual experiment, its average memory cost is 16.2MB, which is the lowest in the three groups of comparative experiments. The FCT\_EDC and AFCT\_GDC methods need to construct a large number of FCTs in the memory, thus causing relatively large memory overhead.

The RTGR of NHSR-OSW using different methods is shown in Table~\ref{rtgr_table} in detail. The FCT\_EDC method tracks all functions of the target software, which greatly increases the response time of the service. Especially for high-performance software, the RTGR of full tracking method on Redis and SQLite are 684\% and 119\%, respectively. The AFCT data model proposed in this paper can adaptive control the tracking overhead. Parameter P sets the percentage of $w_{ub}$ and $w_{total}$ (the total execution frequency of all functions) in \eqref{MWSCP_UBC_contraint}. When P is set to 10\%, AFETM has the highest value of FLR on Nginx (92\%), Httpd (92\%) and SQlite (89\%), and the RTGR on Redis, Nginx, Httpd and SQLite are 81\%, 19\%, 19\% and 49\%, respectively. P=10\% is the default parameter setting in above experiments. Compared with P=10\%, when P=50\% the FLR increases 3\%, but the RTGR increases 98\% in average. Moreover, the RTGR can be further reduced by setting a smaller P according to the performance requirements.

\begin{table}[tb]
\renewcommand{\arraystretch}{1.2}
\caption{Response time growth rate for NHSR-OSW using different methods}
\label{rtgr_table}
\centering
\begin{tabular}{ccccc}
\hline
\multirow{3}{*}{Method} & \multicolumn{4}{c}{Response time growth rate (RTGR)}\\
\cmidrule(lr){2-5}
& \multirow{2}{*}{FCT\_EDC} & \multicolumn{3}{c}{AFCT\_GDC / AFCT\_GCN}\\
\cmidrule(lr){3-5}
&& P=10\% & P=30\% & P=50\% \\
\hline
Redis & 684\% & 81\% & 190\% & 373\% \\
Nginx & 94\% & 19\% & 36\% & 50\% \\
Httpd & 55\% & 19\% & 25\% & 36\% \\
SQlite & 119\% & 49\% & 96\% & 103\% \\
\hline
\end{tabular}
\end{table}

\section{Related Works}
We discuss the related works of this paper from two parts: data collection and data analysis.

\textbf{Data collection:} Existing methods for software system fault diagnosis make use of a variety of input data, such as existing software native log  \cite{Deeplog,SystemlogExp,Draco,SystemlogStr}, casual related events generated by inter-component tracking \cite{target_fault_injection, vPath, Dapper, Magpie, Xtrace, Canopy, Pivot, workflow_centric_tracing} and intra-component tracking \cite{trace_overhead}, system-level metrics and performance counters data \cite{ganglia_distributed_monitoring}, and application-level aggregation data \cite{Moara,Pivot,Astrolabe}.

Many methods record causality between events explicitly by propagating identifiers along the request execution path \cite{target_fault_injection, vPath, Dapper, Magpie, Xtrace, Canopy, Pivot, workflow_centric_tracing}. Prior works collect trace data and propagating identifiers through instrumenting the underlying communication and task management shared lib functions before the system runs \cite{Dapper,Xtrace}. Recent works have also used source and binary rewriting techniques to automatically instrument common execution modules at runtime with out stopping the running system \cite{Pivot,WebPerf,Domino,Timecard}. But these works still select the trace points typically corresponds to some pre-defined events, such as a client sends a request, a low-level IO operation completes, an external RPC is invoked, and so on \cite{target_fault_injection,vPath,Pivot}. Wang et al. proposed a full function execution trace monitoring method to automatically detect faults and locate root causes in function granularity \cite{trace_overhead}. But trace all functions without selection will lead to high tracking overhead. Trace points selection is usually depends on domain knowledge. And even if the historical experience of homogeneous systems can bring some benefits \cite{workflow_centric_tracing}, the lacking of an automatic tracking point selection method still limits the adoption of the function level tracking. 
In our paper, AFETM automatically generates trace points by analyzing the function execution frequency and program structure coverage.

An alternative approach to instrument systems is to infer correlation or causality from existing data such as log data \cite{Canopy}. Approaches include analyzing exist identifiers in logs, such as request ID and IP address, to correlate related log records \cite{Draco,Mining_console_logs,Magpie,The_mystery_machine,Stardust,Non-intrusive_performance_profiling}, and inferring causality using statistical methods \cite{Mining_console_logs,Inferring_models,Modeling_the_Parallel,Using_correlated_surprise} and static source code analysis methods \cite{Mining_console_logs,Non-intrusive_performance_profiling,lprof}. But using existing data have two important limitations: what gets recorded is pre-defined, and the information is recorded without dedicated design which making it hard to correlate events that cross boundaries.

\textbf{Data analysis:} Previous works in performance anomaly analysis and fault diagnosis have presented a variety of automated and semi-automatically analysis methods. Wang et al. provide a comprehensive overview of data center troubleshooting methods in \cite{Performance_troubleshooting}. 

The structural differences between ‘before’ and ‘after’ traces can be obtained by statistical analysis \cite{Comparing_request_flows,Modeling_the_Parallel}. The structure and timing deviations from the normal requests can be found by manual analysis \cite{Comparing_request_flows,Magpie,Performance_debugging,Path-Based_Failure,Pinpoint,Diagnosing_latency}. The slow components or functions can be pinpointed by analyzing response time \cite{Dapper,Whodunit,Modeling_the_Parallel}. The per-workload and per-resource demand information with end-to-end traces of requests can be extracted to analyze performance anomalies \cite{Magpie,Modeling_the_Parallel,Stardust}. Tree structure is the commonly used data model to represent the structure of request execution process \cite{target_fault_injection,trace_overhead}. Pham et al. calculated Gaussian distance between runtime flows and fault flows stored in failure database to help identify root causes of failures \cite{target_fault_injection}. Wang et al. located fault by calculating tree edit distance between runtime FCTs and normal FCTs \cite{trace_overhead}. But the calculation of these distances is time-consuming and will cause a long diagnosis delay. For AFETM, a graph convolutional network and fault injection based method is introduced to solve this problem, and a novel data model AFCT with its construction method is proposed to work with the adaptive tracking.

\section{Conclusion}
Log based fault diagnosis is often inaccurate due to the limitation of the coverage of errors. As a highly flexible runtime software execution process recording technology, dynamic tracking can greatly enrich the fault diagnosis data. However, the high cost of tracking, the amount of up-front effort required to select trace points and the lack of data analysis model limit the adoption of dynamic tracking in fault diagnosis at runtime. This paper presents a fault diagnosis method based on AFETM, including an adaptive tracke point selection method that comprehensively optimizes tracking overhead and structural coverage, an AFCT data model and its construction method, an error characteristic data collection method based on fault injection, and a GCN based fault diagnosis method. Comparative experiments were carried out with three representative methods on NHSR-OSW benchmark composed of Redis, Nginx, Httpd and SQLite. The experimental results show that the proposed method outperforms the existing methods in the accuracy of fault diagnosis, overhead and performance impact on the target system. An important contribution of this method is that it enables the tracking cost to be flexibly controlled according to requirements.



\end{document}